\documentclass[prl,showpacs,showkeys,byrevtex,twocolumn]{revtex4}%
\pdfoutput=1
\usepackage{amsfonts}
\usepackage{amsmath}
\usepackage{amssymb}
\usepackage{graphicx}%
\providecommand{\U}[1]{\protect\rule{.1in}{.1in}}
\newtheorem{theorem}{Theorem}

\newtheorem{definition}{Definition}

\begin{document}
\preprint{ }
\title[ ]{Entanglement-Assisted Quantum Error Correction with Linear Optics}
\author{Mark M. Wilde}
\email{mark.wilde@usc.edu}
\author{Hari Krovi}
\author{Todd A. Brun}
\affiliation{Communication Sciences Institute, Department of Electrical Engineering,
University of Southern California, Los Angeles, California 90089, USA}

\begin{abstract}
We construct a theory of continuous-variable entanglement-assisted quantum
error correction. We present an example of a continuous-variable
entanglement-assisted code that corrects for an arbitrary single-mode error.
We also show how to implement encoding circuits using passive optical devices,
homodyne measurements, feedforward classical communication, conditional
displacements, and off-line squeezers.

\end{abstract}
\volumeyear{2007}
\volumenumber{ }
\issuenumber{ }
\eid{ }
\date{\today}
\received{\today}

\revised{}

\accepted{}

\published{}

\startpage{1}
\endpage{ }
\maketitle

\section{Introduction}

Entanglement is a critical resource for quantum information processing. Shared
entanglement between a sender and receiver enables several quantum
communication protocols such as teleportation \cite{PhysRevLett.70.1895}\ and
superdense coding \cite{PhysRevLett.69.2881}. Brun, Devetak, and Hsieh
exploited the resource of shared entanglement to form a general theory of
quantum error-correcting codes---the entanglement-assisted stabilizer
formalism \cite{science2006brun,arx2006brun}.

Standard quantum error-correcting codes protect a set of qubits from
decoherence by encoding the qubits in a subspace of a larger Hilbert space
\cite{PhysRevA.52.R2493,PhysRevA.54.1098,PhysRevLett.77.793,PhysRevA.54.1862}.
These quantum codes protect a state against a particular error set. Quantum
errors in the error set then either leave the set of qubits invariant or they
take the state out of the subspace into an orthogonal subspace. Measurements
can diagnose which subspace the state is in without disturbing the state. One
can then reverse the effect of the error by rotating the state back into the
original subspace.

Calderbank et al. figured out clever ways of importing classical codes for use
in quantum error correction \cite{ieee1998calderbank}. These methods translate
the classical code to a quantum code. The problem is that the classical codes
have to satisfy a dual-containing constraint. The dual-containing constraint
is equivalent to the operators in the quantum code forming a commuting set.
Few classical codes satisfy the dual-containing constraint so classical theory
was only somewhat useful for quantum error correction after Calderbank et
al.'s results.

Bowen provided the first clue for extending the stabilizer formalism by
constructing an example of a quantum error-correcting code exploiting shared
entanglement \cite{PhysRevA.66.052313}. Brun, Devetak, and Hsieh\ then
established the entanglement-assisted stabilizer formalism
\cite{science2006brun,arx2006brun}.

Entanglement-assisted codes have several key benefits. One can construct an
entanglement-assisted code from an arbitrary linear classical code. The
classical code need not be dual-containing because an entanglement-assisted
code does not require a commuting stabilizer. We turn anticommuting elements
into commuting ones by employing shared entanglement. Thus we can use the
whole of classical coding theory for quantum error correction. Additionally, a
source of pre-established entanglement boosts the rate of an
entanglement-assisted code. The performance of an entanglement-assisted
quantum code follows from that of the imported classical code so that a good
classical code translates to a good quantum code. Entanglement-assisted codes
can also operate in a catalytic manner for quantum computation if a few qubits
are immune to noise \cite{science2006brun,arx2006brun}.

Continuous-variable quantum information has become increasingly popular due to
the practicality of its experimental implementation
\cite{revmod2005braunstein}. Error correction routines are necessary for
proper operation of a continuous-variable quantum communications system.
Braunstein \cite{prl1998braunstein_error}\ and Lloyd and Slotine
\cite{PhysRevLett.80.4088}\ independently proposed the first
continuous-variable\ quantum error-correcting codes. Braunstein's scheme has
the advantage that only linear optical devices and squeezed states prepared
off-line implement the encoding circuit
\cite{prl1998braunstein_error,nat1998braun}. The performance of the code
depends solely on the performance of the off-line squeezers, beamsplitters,
and photodetectors. The disadvantage of Braunstein's scheme is that small
errors accumulate as the computation proceeds if the performance of squeezers
and photodetectors is not sufficient to detect these small errors
\cite{PhysRevA.64.012310}.

\begin{figure}
[ptb]
\begin{center}
\includegraphics[
natheight=4.326700in,
natwidth=8.860000in,
height=1.3482in,
width=3.32in
]%
{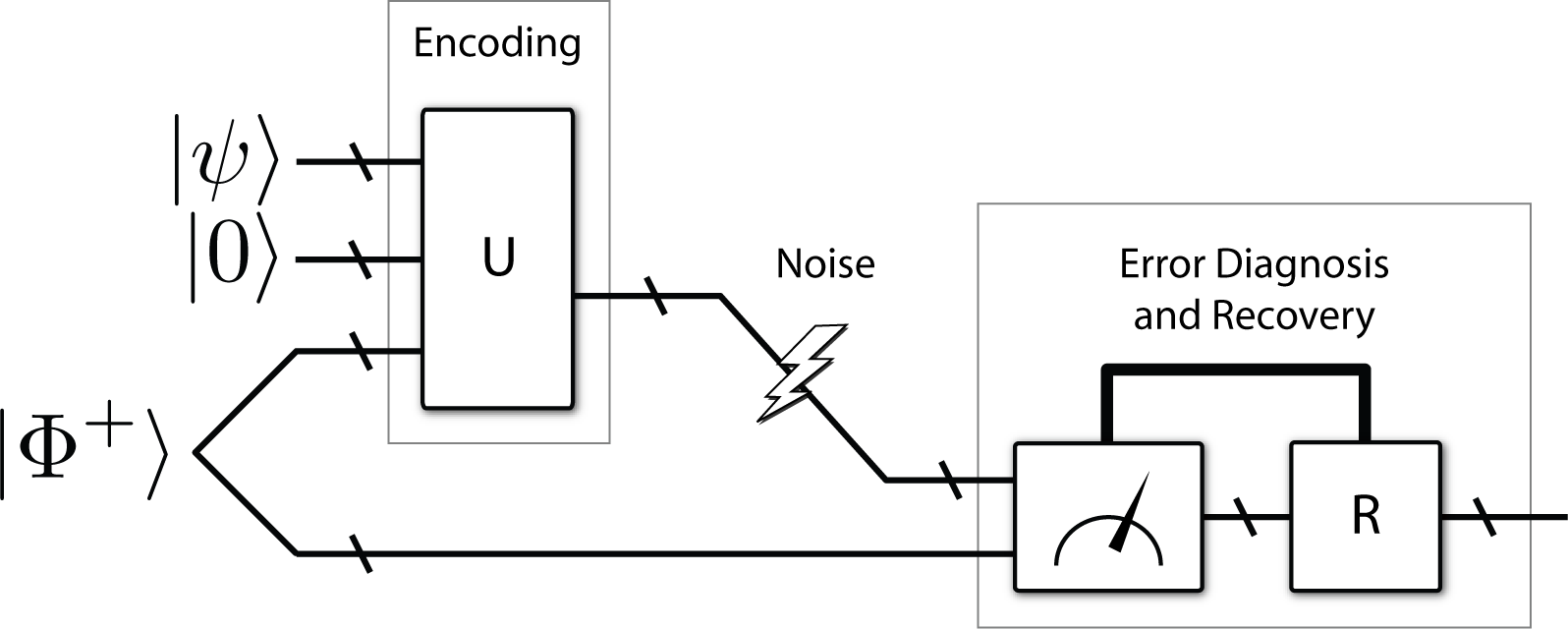}%
\caption{The above figure demonstrates the operation of a continuous-variable
entanglement-assisted code. Lines with bars through them denote multiple
modes. Thin lines denote quantum information and thick lines denote classical
information. Alice possesses states $\left\vert \varphi\right\rangle $,
$\left\vert 0\right\rangle $, and half of the entangled modes $\left\vert
\Phi^{+}\right\rangle $. Bob possesses the other half of entangled modes
$\left\vert \Phi^{+}\right\rangle $. The unitary $U$ encodes the multi-mode
state $\left\vert \varphi\right\rangle $ with the help of several
position-quadrature squeezed ancillas $\left\vert 0\right\rangle $ and
entangled modes $\left\vert \Phi^{+}\right\rangle $. Alice sends her modes
over a noisy quantum channel. The entanglement-assisted communication paradigm
assumes that the noisy channel affects Alice's modes only. Bob measures all
the modes to diagnose the errors and corrects them with a recovery operator
$R$. Bob can perform these measurements with homodyne detection.}%
\label{fig:ea-code}%
\end{center}
\end{figure}

In this paper, we extend the entanglement-assisted stabilizer formalism\ to
continuous-variable\ quantum information \cite{revmod2005braunstein}. Figure
\ref{fig:ea-code} illustrates how a continuous-variable entanglement-assisted
code operates. Brun, Devetak, and Hsieh\ constructed the entanglement-assisted
stabilizer formalism in terms of a sympletic space $\mathbb{Z}_{2}^{2n}$\ over
the field $\mathbb{Z}_{2}$. The theory behind
continuous-variable\ entanglement-assisted quantum error-correcting codes
exploits a symplectic vector space $\mathbb{R}^{2n}$ over the field
$\mathbb{R}$.%

We first review the relation between symplectic spaces, unitary operators, and
the canonical operators for single and multiple modes. We present two theorems
that play a crucial role in constructing continuous-variable
entanglement-assisted codes. We then provide a canonical code and show how a
symplectic transformation relates an arbitrary code to the canonical one. Our
presentation parallels the approach for qubits \cite{arx2006brun}. The
performance of our codes depends solely on the level of squeezing and
photodetector efficiency that is technologically feasible. We give an example
of a continuous-variable entanglement-assisted quantum error-correcting code
that corrects a arbitrary single-mode error.

Our entanglement-assisted quantum error-correcting codes are vulnerable to
finite squeezing effects and inefficient photodetectors for the same reasons
as those given in \cite{prl1998braunstein_error}. Our scheme works well if the
errors due to finite squeezing and inefficiencies in beamsplitters and
photodetectors are smaller than the actual errors.

Our second contribution is an algorithm for constructing the encoding circuit
using linear optics. We refer to any scheme implementing an optical circuit
with passive optical elements, homodyne measurements, feedforward control,
conditional displacements, and off-line squeezers as a linear-optical scheme.
The algorithm exploits and extends previous techniques
\cite{filip:042308,hostens:042315}. The algorithm employs a symplectic
Gaussian elimination technique to decompose an arbitrary encoding circuit into
a linear-optical\ circuit. The transmission amplitudes and phase shifts of
passive beamsplitters encode all the logic rather than the interaction
strength of nonlinear devices.

\section{Symplectic Algebra for Continuous Variables}

We first review some mathematical preliminaries. The notation we develop is
useful for stating Theorems \ref{thm:symp_decomp}\ and \ref{thm_unitary}%
\ precisely. Theorems \ref{thm:symp_decomp}\ and \ref{thm_unitary}\ are
relevant for constructing an entanglement-assisted quantum code and are
analogous to the theorems in \cite{science2006brun,arx2006brun}\ for discrete variables.

We relate the $n$-mode phase-free Heisenberg-Weyl group $\left(  \left[
\mathcal{W}^{n}\right]  ,\ast\right)  $\ to the additive group $\left(
\mathbb{R}^{2n},+\right)  $. Let $X\left(  x\right)  $ be a single-mode
position translation by $x$ and let $Z\left(  p\right)  $ be a single-mode
momentum kick by $p$ where%
\begin{align}
X\left(  x\right)   &  \equiv\exp\left\{  -i\pi x\hat{p}\right\}  ,\nonumber\\
Z\left(  p\right)   &  \equiv\exp\left\{  i\pi p\ \hat{x}\right\}  ,
\end{align}
and $\hat{x}$ and $\hat{p}$ are the position-quadrature and
momentum-quadrature operators respectively. The canonical commutation
relations are $\left[  \hat{x},\hat{p}\right]  =i$. Denote the single-mode
Heisenberg-Weyl group by $\mathcal{W}$ where%
\begin{equation}
\mathcal{W}\equiv\left\{  X\left(  x\right)  Z\left(  p\right)  \ |\ x,p\in
\mathbb{R}\right\}  .
\end{equation}
Let $\mathcal{W}^{n}$ be the set of all $n$-mode operators of the form
$\mathbf{A}\equiv A_{1}\otimes\cdots\otimes A_{n}$ where $A_{j}\in
\mathcal{W\ \ \ }\forall j\in\left\{  1,\ldots,n\right\}  $. Define the
equivalence class%
\begin{equation}
\left[  \mathbf{A}\right]  \equiv\left\{  \beta\mathbf{A}\ |\ \beta
\in\mathbb{C},\left\vert \beta\right\vert =1\right\}
\end{equation}
with representative operator having $\beta=1$. The above equivalence class is
useful because global phases are not relevant in the formulation of our codes.
The group operation $\ast$ for the above equivalence class is as follows%
\begin{align}
\left[  \mathbf{A}\right]  \ast\left[  \mathbf{B}\right]   &  \equiv\left[
A_{1}\right]  \ast\left[  B_{1}\right]  \otimes\cdots\otimes\left[
A_{n}\right]  \ast\left[  B_{n}\right]  \nonumber\\
&  =\left[  A_{1}B_{1}\right]  \otimes\cdots\otimes\left[  A_{n}B_{n}\right]
=\left[  \mathbf{AB}\right]  .
\end{align}
The equivalence class $\left[  \mathcal{W}^{n}\right]  =\left\{  \left[
\mathbf{A}\right]  :\mathbf{A}\in\mathcal{W}^{n}\right\}  $ forms a
commutative group $\left(  \left[  \mathcal{W}^{n}\right]  ,\ast\right)  $. We
name $\left(  \left[  \mathcal{W}^{n}\right]  ,\ast\right)  $ the
\textit{phase-free Heisenberg-Weyl group}.

Consider the $2n$-dimensional real vector space $\mathbb{R}^{2n}$. It forms
the commutative group $\left(  \mathbb{R}^{2n},+\right)  $ with operation $+$
defined as vector addition. We employ the notation $\mathbf{u}=\left(
\mathbf{p}|\mathbf{x}\right)  ,\mathbf{v}=\left(  \mathbf{p}^{\prime
}|\mathbf{x}^{\prime}\right)  $ to represent any vectors $\mathbf{u,v}%
\in\mathbb{R}^{2n}$ respectively. Each vector $\mathbf{p}$ and $\mathbf{x}$
has elements $\left(  p_{1},\ldots,p_{n}\right)  $ and $\left(  x_{1}%
,\ldots,x_{n}\right)  $ respectively with similar representations for
$\mathbf{p}^{\prime}$ and $\mathbf{x}^{\prime}$. The \textit{symplectic
product} $\odot$ of $\mathbf{u}$ and $\mathbf{v}$ is%
\begin{equation}
\mathbf{u}\odot\mathbf{v\equiv p\cdot x}^{\prime}-\mathbf{x\cdot p}^{\prime
}=\sum_{i=1}^{n}p_{i}x_{i}^{\prime}-x_{i}p_{i}^{\prime},
\end{equation}
where $\cdot$ is the standard inner product. Define a map $\mathbf{D}%
:\mathbb{R}^{2n}\rightarrow\mathcal{W}^{n}$ as follows:%
\begin{equation}
\mathbf{D}\left(  \mathbf{u}\right)  \equiv\exp\left\{  i\sqrt{\pi}%
{\textstyle\sum\limits_{i=1}^{n}}
\left(  p_{i}\hat{x}_{i}-x_{i}\hat{p}_{i}\right)  \right\}
.\label{eq:map_symp_hw}%
\end{equation}
Let%
\begin{align}
\mathbf{X}\left(  \mathbf{x}\right)   &  \equiv X\left(  x_{1}\right)
\otimes\cdots\otimes X\left(  x_{n}\right)  ,\nonumber\\
\mathbf{Z}\left(  \mathbf{p}\right)   &  \equiv Z\left(  p_{1}\right)
\otimes\cdots\otimes Z\left(  p_{n}\right)  ,
\end{align}
so that $\mathbf{D}\left(  \mathbf{p|x}\right)  $ and $\mathbf{Z}\left(
\mathbf{p}\right)  \mathbf{X}\left(  \mathbf{x}\right)  $ belong to the same
equivalence class:%
\begin{equation}
\left[  \mathbf{D}\left(  \mathbf{p|x}\right)  \right]  =\left[
\mathbf{Z}\left(  \mathbf{p}\right)  \mathbf{X}\left(  \mathbf{x}\right)
\right]  .
\end{equation}
The map $\left[  \mathbf{D}\right]  :\mathbb{R}^{2n}\rightarrow\left[
\mathcal{W}^{n}\right]  $ is an isomorphism%
\begin{equation}
\left[  \mathbf{D}\left(  \mathbf{u+v}\right)  \right]  =\left[
\mathbf{D}\left(  \mathbf{u}\right)  \right]  \left[  \mathbf{D}\left(
\mathbf{v}\right)  \right]  ,
\end{equation}
where $\mathbf{u,v}\in\mathbb{R}^{2n}$. We use the BCH\ theorem $e^{A}%
e^{B}=e^{B}e^{A}e^{\left[  A,B\right]  }$\ and the symplectic product to
capture the commutation relations of any operators $\mathbf{D}\left(
\mathbf{u}\right)  $ and $\mathbf{D}\left(  \mathbf{v}\right)  $:%
\begin{equation}
\mathbf{D\left(  \mathbf{u}\right)  D}\left(  \mathbf{v}\right)  =\exp\left\{
i\pi\left(  \mathbf{u}\odot\mathbf{v}\right)  \right\}  \mathbf{D}\left(
\mathbf{v}\right)  \mathbf{D}\left(  \mathbf{u}\right)  .
\end{equation}
The operators $\mathbf{D\left(  \mathbf{u}\right)  }$ and $\mathbf{D\left(
\mathbf{v}\right)  }$ commute if $\mathbf{u}\odot\mathbf{v}=2n$ and
anticommute if $\mathbf{u}\odot\mathbf{v}=2n+1$ for any $n\in\mathbb{Z}$. The
set of canonical operators $\hat{x}_{i},\hat{p}_{i}$ for all $i\in\left\{
1,\ldots,n\right\}  $ have the canonical commutation relations:%
\begin{align*}
\left[  \hat{x}_{i},\hat{x}_{j}\right]   &  =0,\\
\left[  \hat{p}_{i},\hat{p}_{j}\right]   &  =0,\\
\left[  \hat{x}_{i},\hat{p}_{j}\right]   &  =i\delta_{ij}.
\end{align*}
Let $\mathcal{T}^{n}$ be the set of all linear combinations of the canonical
operators:%
\begin{equation}
\mathcal{T}^{n}\equiv\left\{  \sum_{i=1}^{n}\alpha_{i}\hat{x}_{i}+\beta
_{i}\hat{p}_{i}:\forall i,\ \ \alpha_{i},\beta_{i}\in\mathbb{R}\right\}  .
\end{equation}
Define the map $\mathbf{M}:\mathbb{R}^{2n}\rightarrow\mathcal{T}^{n}$ as%
\begin{equation}
\mathbf{M}\left(  \mathbf{u}\right)  \equiv\mathbf{u\cdot\hat{R}}%
^{n},\label{eq:map_M}%
\end{equation}
where $\mathbf{u}=\left(  \mathbf{p}|\mathbf{x}\right)  \in\mathbb{R}^{2n}$,%
\begin{equation}
\mathbf{\hat{R}}^{n}=\left[  \left.
\begin{array}
[c]{ccc}%
\hat{x}_{1} & \cdots & \hat{x}_{n}%
\end{array}
\right\vert
\begin{array}
[c]{ccc}%
\hat{p}_{1} & \cdots & \hat{p}_{n}%
\end{array}
\right]  ^{T},
\end{equation}
and $\cdot$ is the inner product. We can now write $\mathcal{T}^{n}%
\equiv\left\{  \mathbf{M}\left(  \mathbf{u}\right)  :\mathbf{u}\in
\mathbb{R}^{2n}\right\}  $. The symplectic product gives the commutation
relations of elements of $\mathcal{T}^{n}$:%
\begin{equation}
\left[  \mathbf{M}\left(  \mathbf{u}\right)  ,\mathbf{M}\left(  \mathbf{v}%
\right)  \right]  =\left(  \mathbf{u}\odot\mathbf{v}\right)  i.
\end{equation}
The definitions given below provide terminology used in the statements of
Theorems \ref{thm:symp_decomp}\ and \ref{thm_unitary} and used in the
construction of our continuous-variable entanglement-assisted codes.

\begin{definition}
A subspace $V$ of a space $W$ is \textit{symplectic} if there is no
$\mathbf{v}\in V$ such that $\forall\ \mathbf{u}\in V:\mathbf{u}%
\odot\mathbf{v}=0$.
\end{definition}

\begin{definition}
A subspace $V$ of a space $W$ is \textit{isotropic} if $\forall\ \mathbf{u}\in
W\mathbf{,v}\in V:\mathbf{u}\odot\mathbf{v}=0$.
\end{definition}

\begin{definition}
Two vectors $\mathbf{u},\mathbf{v}\in\mathbb{R}^{2n}$ form a
\textit{hyperbolic pair} $\left(  \mathbf{u},\mathbf{v}\right)  $ if
$\mathbf{u}\odot\mathbf{v}=1$.
\end{definition}

\begin{definition}
The \textit{symplectic dual} $V^{\perp}$\ of a subspace $V$ is $V^{\bot}%
\equiv\left\{  \mathbf{w}:\mathbf{w}\odot\mathbf{u}=0,\ \ \forall
\ \mathbf{u}\in V\right\}  $.
\end{definition}

\begin{definition}
A \textit{symplectic matrix} $\mathbf{\Upsilon}:\mathbb{R}^{2n}\rightarrow
\mathbb{R}^{2n}$ preserves the symplectic product:%
\begin{equation}
\mathbf{\Upsilon u}\odot\mathbf{\Upsilon v}=\mathbf{u}\odot
\mathbf{v\ \ \ \ \ \ \ \ \ }\forall\ \mathbf{u},\mathbf{v}\in\mathbb{R}^{2n}.
\end{equation}
It satisfies the condition $\mathbf{\Upsilon}^{T}\mathbf{J\Upsilon}%
=\mathbf{J}$ where%
\begin{equation}
\mathbf{J}=%
\begin{bmatrix}
\mathbf{0}_{n\times n} & \mathbf{I}_{n\times n}\\
-\mathbf{I}_{n\times n} & \mathbf{0}_{n\times n}%
\end{bmatrix}
.
\end{equation}

\end{definition}

\section{Theorems for Entanglement-Assisted Quantum Error Correction for
Continuous-Variable Systems}

Theorem \ref{thm:symp_decomp} applies to parity check matrices for our
continuous-variable\ entanglement-assisted codes. The theorem gives an optimal
way of decomposing an arbitrary subspace of $\mathbb{R}^{2n}$ into a purely
isotropic subspace and a purely symplectic subspace. Thus we can decompose the
rows of an arbitrary parity check matrix in this fashion. We later see that
this theorem determines how much entanglement is necessary for the code.

\begin{theorem}
\label{thm:symp_decomp}Let $V$ be a subspace of $\mathbb{R}^{2n}$. Suppose
$\dim\left(  V\right)  =m$. There exists a symplectic subspace $\mathrm{symp}%
\left(  V\right)  =\ \mathrm{span}\left\{  \mathbf{u}_{1},\ldots
,\mathbf{u}_{c},\mathbf{v}_{1},\ldots,\mathbf{v}_{c}\right\}  $ of
$\mathbb{R}^{2n}$ where $\dim\left(  \mathrm{symp}\left(  V\right)  \right)
=2c$. The hyperbolic pairs $\left(  \mathbf{u}_{i},\mathbf{v}_{i}\right)  $
where $i=1,\ldots,c$ span $\mathrm{symp}\left(  V\right)  $. There exists an
isotropic subspace $\mathrm{iso}\left(  V\right)  =\ \mathrm{span}\left\{
\mathbf{u}_{c+1},\ldots,\mathbf{u}_{c+l}\right\}  $ where $\dim\left(
\mathrm{iso}\left(  V\right)  \right)  =l$. Subspace $V$ has dimension
$m=2c+l$ and is the direct sum of its isotropic and symplectic subspaces:
$V=\ \mathrm{iso}\left(  V\right)  \oplus\ \mathrm{symp}\left(  V\right)  $.
\end{theorem}

A constructive proof of the above theorem is in \cite{book2001symp}. The set
of basis vectors for $\mathrm{iso}\left(  V\right)  $ corresponds to a
commuting set of\ observables in both $\mathcal{W}^{n}$ and $\mathcal{T}^{n}$
using the maps $\mathbf{D}$\ and $\mathbf{M}$ respectively. Each hyperbolic
pair $\left(  \mathbf{u}_{i},\mathbf{v}_{i}\right)  $ in $\mathrm{symp}\left(
V\right)  $\ corresponds\ via $\mathbf{D}$ to a pair of observables in
$\mathcal{W}^{n}$\ that anticommute and corresponds via $\mathbf{M}$ to a pair
in $\mathcal{T}^{n}$ with commutator $\left[  \mathbf{M}\left(  \mathbf{u}%
_{i}\right)  ,\mathbf{M}\left(  \mathbf{v}_{i}\right)  \right]  =i$.

Theorem \ref{thm_unitary} is useful in relating a general
continuous-variable\ entanglement-assisted quantum error-correcting code\ to a
canonical one (described below) by a unitary operator. The unitary operator
corresponds to an encoding circuit for the code.

\begin{theorem}
\label{thm_unitary}There exists a unitary operator $U_{\mathbf{\Upsilon}}$
corresponding to a symplectic matrix $\Upsilon$ so that the following two
conditions hold $\forall\ \mathbf{u}\in R^{2n}$:%
\begin{align}
\left[  \mathbf{D}\left(  \mathbf{\Upsilon u}\right)  \right]   &  =\left[
U_{\mathbf{\Upsilon}}\mathbf{D}\left(  \mathbf{u}\right)  U_{\mathbf{\Upsilon
}}^{-1}\right]  ,\nonumber\\
\mathbf{M}\left(  \mathbf{\Upsilon u}\right)   &  =U_{\mathbf{\Upsilon}%
}\mathbf{M}\left(  \mathbf{u}\right)  U_{\mathbf{\Upsilon}}^{-1}.
\end{align}

\end{theorem}

Theorem \ref{thm_unitary}\ is a consequence of the Stone-von Neumann theorem
\cite{eisert-2003-1}.

The unitary $U_{\mathbf{\Upsilon}}^{-1}$\ for the encoding circuit relates a
general continuous-variable\ entanglement-assisted quantum error-correcting
code\ to the canonical one.

\section{Canonical Entanglement-Assisted Quantum error-correcting Code}

We first consider a code protecting against a canonical error set
$S_{0}\subset\mathbb{R}^{2n}$ with errors $\mathbf{D}\left(  \mathbf{u}%
\right)  $ where $\mathbf{u}\in\mathbb{R}^{2n}$. We later extend to a more
general error set by applying Theorem~2.

Continuous-variable\ errors are equivalent to translations in position and
kicks in momentum \cite{prl1998braunstein_error,PhysRevA.64.012310}. These
errors correspond to vectors in $\mathbb{R}^{2n}$ via the inverse map
$\mathbf{D}^{-1}$.

Suppose Alice wishes to protect a $k$-mode quantum state $\left\vert
\varphi\right\rangle $:%
\begin{equation}
\left\vert \varphi\right\rangle =%
{\textstyle\idotsint}
dx_{1}\cdots dx_{k}\ \varphi\left(  x_{1},\ldots,x_{k}\right)  \ \left\vert
x_{1}\right\rangle \cdots\left\vert x_{k}\right\rangle .
\end{equation}
Alice and Bob possess $c$ sets of infinitely-squeezed, perfectly entangled
states $\left\vert \Phi\right\rangle ^{\otimes c}$ where%
\[
\left\vert \Phi\right\rangle \equiv\left(  \int\ dx\ \left\vert x\right\rangle
\left\vert x\right\rangle \right)  /\sqrt{\pi}.
\]
The state $\left\vert \Phi\right\rangle $ is a zero-valued eigenstate of the
relative position observable $\hat{x}_{A}-\hat{x}_{B}$ and total momentum
observable $\hat{p}_{A}+\hat{p}_{B}$. Alice possesses $l=n-k-c$ ancilla
registers initialized to infinitely-squeezed zero-position eigenstates of the
position observables $\hat{x}_{k+1},\ldots,\hat{x}_{k+l}$: $\left\vert
\mathbf{0}\right\rangle =\left\vert 0\right\rangle ^{\otimes l}$. She encodes
the state $\left\vert \varphi\right\rangle $ with the canonical isometric
encoder $U_{0}$\ as follows:%
\begin{equation}
U_{0}:\left\vert \varphi\right\rangle \left\vert \Phi\right\rangle ^{\otimes
c}\rightarrow\left\vert \varphi\right\rangle \left\vert \mathbf{0}%
\right\rangle \left\vert \Phi\right\rangle ^{\otimes c}.
\label{eq:canon_encode}%
\end{equation}
The canonical code corrects the error set%
\begin{equation}
S_{0}=\left\{
\begin{array}
[c]{c}%
\left(  \alpha\left(  \mathbf{a},\mathbf{a}_{1},\mathbf{a}_{2}\right)
,\mathbf{b,a}_{2}|\beta\left(  \mathbf{a},\mathbf{a}_{1},\mathbf{a}%
_{2}\right)  ,\mathbf{a},\mathbf{a}_{1}\right) \\
:\mathbf{b,a}\in\mathbb{R}^{l},\mathbf{a}_{1}\mathbf{,a}_{2}\in\mathbb{R}^{c}%
\end{array}
\right\}  ,
\end{equation}
for some known functions $\alpha,\beta:\mathbb{R}^{l}\times\mathbb{R}%
^{c}\times\mathbb{R}^{c}\rightarrow\mathbb{R}^{k}$. Suppose an error
$\mathbf{D}\left(  \mathbf{u}\right)  $ occurs where%
\begin{equation}
\mathbf{u}=\left(  \alpha\left(  \mathbf{a},\mathbf{a}_{1},\mathbf{a}%
_{2}\right)  ,\mathbf{b,a}_{2}|\beta\left(  \mathbf{a},\mathbf{a}%
_{1},\mathbf{a}_{2}\right)  ,\mathbf{a},\mathbf{a}_{1}\right)  .
\label{eq:cv_error}%
\end{equation}
The state $\left\vert \varphi\right\rangle \left\vert \mathbf{0}\right\rangle
\left\vert \Phi\right\rangle ^{\otimes c}$ becomes (up to a global phase)%
\begin{equation}
\mathbf{Z}\left(  \alpha\right)  \mathbf{X}\left(  \beta\right)  \left\vert
\varphi\right\rangle \otimes\left\vert \mathbf{a}\right\rangle \otimes
\left\vert \mathbf{a}_{1},\mathbf{a}_{2}\right\rangle ,
\end{equation}
where $\left\vert \mathbf{a}\right\rangle =\mathbf{X}\left(  \mathbf{a}%
\right)  \left\vert \mathbf{0}\right\rangle $ and $\left\vert \mathbf{a}%
_{1},\mathbf{a}_{2}\right\rangle =\mathbf{X}\left(  \mathbf{a}_{1}\right)
\mathbf{Z}\left(  \mathbf{a}_{2}\right)  \left\vert \Phi\right\rangle
^{\otimes c}$. Bob measures the position observables of the ancillas
$\left\vert \mathbf{a}\right\rangle $ and the relative position and total
momentum observables of the state $\left\vert \mathbf{a}_{1},\mathbf{a}%
_{2}\right\rangle $. He obtains the reduced error syndrome $\mathbf{r}=\left(
\mathbf{a,a}_{1},\mathbf{a}_{2}\right)  $. The reduced error syndrome
specifies the error up to an irrelevant value of $\mathbf{b}$ in
(\ref{eq:cv_error}). Bob reverses the error $\mathbf{u}$ by applying the map
$\mathbf{D}\left(  -\mathbf{u}^{\prime}\right)  $ where%
\begin{equation}
\mathbf{u}^{\prime}=\left(  \alpha\left(  \mathbf{a},\mathbf{a}_{1}%
,\mathbf{a}_{2}\right)  ,\mathbf{0,a}_{2}|\beta\left(  \mathbf{a}%
,\mathbf{a}_{1},\mathbf{a}_{2}\right)  ,\mathbf{a},\mathbf{a}_{1}\right)  .
\label{eq:reverse_error}%
\end{equation}
The canonical code is degenerate because the $\mathbf{Z}\left(  \mathbf{b}%
\right)  $ errors do not affect the encoded state and Bob does not need to
know $\mathbf{b}$ to correct the errors.

We can describe the operation of the canonical code using binary matrix
algebra. This technique gives a correspondence between the canonical code and
classical coding theory. The following parity check matrix $F$\ characterizes
the errors that the canonical code can correct:%
\begin{equation}
F\equiv\left[  \left.
\begin{array}
[c]{ccc}%
\mathbf{0}_{l\times k} & \mathbf{I}_{l\times l} & \mathbf{0}_{l\times c}\\
\mathbf{0}_{c\times k} & \mathbf{0}_{c\times l} & \mathbf{I}_{c\times c}\\
\mathbf{0}_{c\times k} & \mathbf{0}_{c\times l} & \mathbf{0}_{c\times c}%
\end{array}
\right\vert
\begin{array}
[c]{ccc}%
\mathbf{0}_{l\times k} & \mathbf{0}_{l\times l} & \mathbf{0}_{l\times c}\\
\mathbf{0}_{c\times k} & \mathbf{0}_{c\times l} & \mathbf{0}_{c\times c}\\
\mathbf{0}_{c\times k} & \mathbf{0}_{c\times l} & \mathbf{I}_{c\times c}%
\end{array}
\right]  .\label{eq:bad_parity_check}%
\end{equation}
The rows in the above matrix $F$ correspond to observables via the map
$\mathbf{M}$ in (\ref{eq:map_M}). Bob can measure these observables to
diagnose the error. However, a problem exists. Suppose Bob naively attempts to
learn the error by measuring the observables $\mathbf{M}\left(  \mathbf{f}%
\right)  $ for all rows $\mathbf{f}$\ in $F$. Bob disturbs the state because
these observables do not commute. We remedy this situation later\ by supposing
that Alice and Bob share entanglement as in the above construction in
(\ref{eq:canon_encode}).

Let us define the \textit{canonical symplectic code} $C_{0}$ corresponding to
$F$ to be all the real vectors symplectically orthogonal to the rows of $F$:%
\begin{equation}
C_{0}\equiv\ \text{rowspace}\left(  F\right)  ^{\bot}.
\end{equation}
Let $S_{0}$ be the set of correctable errors. All pairs of errors in $S_{0}$
obey one of the following constraints: $\forall\ \mathbf{u},\mathbf{u}%
^{\prime}\in S_{0}$ with $\mathbf{u}\neq\mathbf{u}^{\prime}$ either
$\mathbf{u}-\mathbf{u}^{\prime}\notin C_{0}$ or $\mathbf{u}-\mathbf{u}%
^{\prime}\in$ iso$\left(  C_{0}^{\bot}\right)  $. The condition $\mathbf{u}%
-\mathbf{u}^{\prime}\notin C_{0}$\ states that an error is correctable if it
has a unique error syndrome. The latter condition applies if any two errors
have the same effect on the encoded state.

The rowspace of $F$ is a $\left(  2c+l\right)  $-dimensional subspace of
$\mathbb{R}^{2n}$. Therefore it decomposes as a direct sum of an isotropic and
symplectic subspace according to Theorem~\ref{thm:symp_decomp}. The first $l$
rows of $F$ are a basis for the isotropic subspace and the last $2c$ rows are
a basis for the symplectic subspace.

We can remedy the problems with the parity check matrix in
(\ref{eq:bad_parity_check}) by constructing an augmented parity check matrix
$F_{aug}$ as%
\[
\left[  \left.
\begin{array}
[c]{cccc}%
\mathbf{0}_{l\times k} & \mathbf{I}_{l\times l} & \mathbf{0}_{l\times c} &
\mathbf{0}_{l\times c}\\
\mathbf{0}_{c\times k} & \mathbf{0}_{c\times l} & \mathbf{I}_{c\times c} &
-\mathbf{I}_{c\times c}\\
\mathbf{0}_{c\times k} & \mathbf{0}_{c\times l} & \mathbf{0}_{c\times c} &
\mathbf{0}_{c\times c}%
\end{array}
\right\vert
\begin{array}
[c]{cccc}%
\mathbf{0}_{l\times k} & \mathbf{0}_{l\times l} & \mathbf{0}_{l\times c} &
\mathbf{0}_{l\times c}\\
\mathbf{0}_{c\times k} & \mathbf{0}_{c\times l} & \mathbf{0}_{c\times c} &
\mathbf{0}_{c\times c}\\
\mathbf{0}_{c\times k} & \mathbf{0}_{c\times l} & \mathbf{I}_{c\times c} &
\mathbf{I}_{c\times c}%
\end{array}
\right]  .
\]
The error-correcting properties of the code are the same as before. The extra
entries correspond to Bob's half of entangled modes shared with Alice. These
extra modes are noiseless because they are on the receiving end of the
channel. The isotropic subspace of rowspace$\left(  F\right)  $ remains the
same in the above construction. The symplectic subspace of rowspace$\left(
F\right)  $ becomes isotropic in the higher dimensional space rowspace$\left(
F_{aug}\right)  $. Each row $\mathbf{f}$\ of $F_{aug}$ corresponds to an
element of the set%
\begin{equation}
\mathcal{M}_{0}\equiv\left\{  \mathbf{M}\left(  \mathbf{f}\right)
:\mathbf{f}\text{ is a row of }F_{aug}\right\}  .
\end{equation}
Observables in $\mathcal{M}_{0}$ commute because rowspace$\left(
F_{aug}\right)  $\ is purely isotropic. Bob can then measure these observables
to learn the error without disturbing the state. The \textit{canonical
codespace} $\mathcal{C}_{0}$ is the simultaneous zero eigenspace of operators
in $\mathcal{M}_{0}$---the encoding in (\ref{eq:canon_encode}) satisfies this
constraint. Measurement of the observables corresponding to the first $l$ rows
of $F_{aug}$\ gives Bob the error vector $\mathbf{a}$. The next $c$
measurements give Bob the error vector $\mathbf{a}_{1}$ and the last $c$
measurements give Bob the error vector $\mathbf{a}_{2}$. This reduced syndrome
$\left(  \mathbf{a,a}_{1},\mathbf{a}_{2}\right)  $ specifies the error up to
an irrelevant value of $\mathbf{b}$. Bob can reverse the error $\mathbf{u}$ by
applying the map $\mathbf{D}\left(  -\mathbf{u}^{\prime}\right)  $ with
$\mathbf{u}^{\prime}$ defined in (\ref{eq:reverse_error}). The number of
entangled modes used in the code is%
\[
c=\dim\left(  \text{iso}\left(  \text{rowspace}\left(  F\right)  \right)
\right)  /2,
\]
and the number of encoded modes is%
\[
k=n-\dim\left(  \text{symp}\left(  \text{rowspace}\left(  F\right)  \right)
\right)  -c.
\]
Thus Alice and Bob can use the above canonical code with entanglement
assistance to correct for a canonical error set.

\section{General Entanglement-Assisted Quantum error-correcting Codes}

We now show how to construct an entanglement-assisted quantum error-correcting
code from an arbitrary subspace $C$ of $\mathbb{R}^{2n}$. We give an example
of this construction as we develop the theory. Suppose that subspace $C$ is
$\left(  2n-m\right)  $-dimensional where $m=2c+l$ for some $c,l\geq0$ and
$c+l<n$. Think of subspace $C$\ as an arbitrary symplectic code. We can find a
symplectic basis $\left\{  \mathbf{u}_{i},\mathbf{v}_{i}\right\}  _{i=1}^{n}%
$\ for $\mathbb{R}^{2n}$ by Theorem~\ref{thm:symp_decomp} with the following
two constraints. First, it has hyperbolic pairs $\left(  \mathbf{u}%
_{i},\mathbf{v}_{i}\right)  $ $i=1,\ldots,n$. Second, $2n-m$ vectors in
$\left\{  \mathbf{u}_{i},\mathbf{v}_{i}\right\}  _{i=1}^{n}$\ correspond to a
basis for $C$ and the other $m$ vectors are a basis for the $m$-dimensional
subspace $C^{\bot}$. Let us define the set%
\begin{equation}
\mathcal{R}\equiv\left\{  \mathbf{u}_{1},\ldots,\mathbf{u}_{c+l}%
,\mathbf{v}_{1},\ldots,\mathbf{v}_{c}\right\}
\end{equation}
as a basis for the $m$-dimensional subspace $C^{\bot}$. Define the set%
\begin{equation}
\mathcal{R}_{0}\equiv\left\{  \mathbf{e}_{1},\ldots,\mathbf{e}_{c+l}%
,\mathbf{e}_{n+1},\ldots,\mathbf{e}_{n+c}\right\}
\end{equation}
as a basis for the canonical subspace $C_{0}^{\bot}$.

How do we find the symplectic basis for $\mathbb{R}^{2n}$? We can employ a
symplectic Gram-Schmidt orthogonalization procedure similar to that outlined
in Ref. \cite{arx2006brun}. Suppose we have an initial arbitrary set of
vectors that form a basis for $C$. We can multiply and add the vectors
together without changing the error-correcting properties of the eventual code
that we formulate. These operations are \textquotedblleft row
operations.\textquotedblright\ Row operations are useful for determining an
alternate set of vectors that determine a basis for $C^{\perp}$. This
alternate set then decomposes into purely symplectic and purely isotropic
parts. 

We turn to an example to highlight the above theory. Consider the following
four vectors:%
\begin{align}
&  \left(  \left.
\begin{array}
[c]{cccc}%
1 & 0 & 1 & 0
\end{array}
\right\vert
\begin{array}
[c]{cccc}%
0 & 1 & 0 & 0
\end{array}
\right)  ,\nonumber\\
&  \left(  \left.
\begin{array}
[c]{cccc}%
1 & 1 & 0 & 1
\end{array}
\right\vert
\begin{array}
[c]{cccc}%
0 & 0 & 0 & 0
\end{array}
\right)  ,\nonumber\\
&  \left(  \left.
\begin{array}
[c]{cccc}%
0 & 1 & 0 & 0
\end{array}
\right\vert
\begin{array}
[c]{cccc}%
1 & 1 & 1 & 0
\end{array}
\right)  ,\nonumber\\
&  \left(  \left.
\begin{array}
[c]{cccc}%
0 & 0 & 0 & 0
\end{array}
\right\vert
\begin{array}
[c]{cccc}%
1 & 1 & 0 & 1
\end{array}
\right)  .
\end{align}
Suppose they span the dual $C^{\perp}$\ of an arbitrary subspace $C$.
$C^{\perp}$\ is then a four-dimensional vector space. This subspace is similar
to one for a discrete-variable\ entanglement-assisted quantum error-correcting
code \cite{science2006brun}. We use it to develop a continuous-variable
entanglement-assisted code. We perform row operations on the above set of
vectors and obtain the following four vectors:%
\begin{align}
\mathbf{u}_{1} &  =\left(  \left.
\begin{array}
[c]{cccc}%
1 & 1 & 0 & 1
\end{array}
\right\vert
\begin{array}
[c]{cccc}%
0 & 0 & 0 & 0
\end{array}
\right)  ,\nonumber\\
\mathbf{u}_{2} &  =\left(  \left.
\begin{array}
[c]{cccc}%
-\sqrt{\frac{1}{2}} & \sqrt{2} & -\sqrt{2} & \sqrt{\frac{1}{2}}%
\end{array}
\right\vert
\begin{array}
[c]{cccc}%
\sqrt{\frac{1}{2}} & -\sqrt{\frac{1}{2}} & \sqrt{\frac{1}{2}} & 0
\end{array}
\right)  ,\nonumber\\
\mathbf{v}_{1} &  =\left(  \left.
\begin{array}
[c]{cccc}%
1 & 0 & 1 & 0
\end{array}
\right\vert
\begin{array}
[c]{cccc}%
0 & 1 & 0 & 0
\end{array}
\right)  ,\nonumber\\
\mathbf{v}_{2} &  =\left(  \left.
\begin{array}
[c]{cccc}%
-\sqrt{2} & \sqrt{\frac{1}{2}} & -\sqrt{\frac{9}{2}} & \sqrt{\frac{1}{2}}%
\end{array}
\right\vert
\begin{array}
[c]{cccc}%
\sqrt{\frac{1}{2}} & -\sqrt{2} & 0 & \sqrt{\frac{1}{2}}%
\end{array}
\right)  .\label{eq:ent-code-vecs}%
\end{align}
The above vectors define a symplectic basis for $C^{\perp}$ and are in the set
$\mathcal{R}$. The above vectors have the same symplectic relations as the
following four standard basis vectors:%
\begin{align}
\mathbf{e}_{1} &  =\left(  \left.
\begin{array}
[c]{cccc}%
1 & 0 & 0 & 0
\end{array}
\right\vert
\begin{array}
[c]{cccc}%
0 & 0 & 0 & 0
\end{array}
\right)  ,\nonumber\\
\mathbf{e}_{2} &  =\left(  \left.
\begin{array}
[c]{cccc}%
0 & 1 & 0 & 0
\end{array}
\right\vert
\begin{array}
[c]{cccc}%
0 & 0 & 0 & 0
\end{array}
\right)  ,\nonumber\\
\mathbf{e}_{5} &  =\left(  \left.
\begin{array}
[c]{cccc}%
0 & 0 & 0 & 0
\end{array}
\right\vert
\begin{array}
[c]{cccc}%
1 & 0 & 0 & 0
\end{array}
\right)  ,\nonumber\\
\mathbf{e}_{6} &  =\left(  \left.
\begin{array}
[c]{cccc}%
0 & 0 & 0 & 0
\end{array}
\right\vert
\begin{array}
[c]{cccc}%
0 & 1 & 0 & 0
\end{array}
\right)  .\label{eq:standard-vecs}%
\end{align}
The above standard basis vectors are in the set $\mathcal{R}_{0}$.

We return to the general theory. A symplectic matrix $\mathbf{\Upsilon}$ then
exists that maps the hyperbolic pairs $\left(  \mathbf{u}_{i},\mathbf{v}%
_{i}\right)  $ to the standard hyperbolic pairs $\left(  \mathbf{e}%
_{i},\mathbf{e}_{n+i}\right)  $ for all $i$ \cite{book2001symp}. Let $H$ and
$F$ be the matrices whose rows consist of elements of $\mathcal{R}$ and
$\mathcal{R}_{0}$ respectively. Let $H_{\text{aug}}$ and $F_{\text{aug}}$ be
the augmented versions of $H$ and $F$ respectively. Then $H\ \mathbf{\Upsilon
}^{T}=F$ and $H_{\text{aug}}P\mathbf{\Upsilon}^{T}P^{T}=F_{\text{aug}}$ where
$P$ is a permutation matrix that makes columns $n+1$ through $n+c$ be the last
$c$ columns and shifts columns $n+c+1$ through $2n+c$ left by $c$ positions.

The four vectors in (\ref{eq:standard-vecs}) determine a canonical
entanglement-assisted code. We place them as row vectors in a parity check
matrix $F$:%
\begin{equation}
F=\left[
\begin{array}
[c]{cccc}%
1 & 0 & 0 & 0\\
0 & 1 & 0 & 0\\
0 & 0 & 0 & 0\\
0 & 0 & 0 & 0
\end{array}
\right\vert \left.
\begin{array}
[c]{cccc}%
0 & 0 & 0 & 0\\
0 & 0 & 0 & 0\\
1 & 0 & 0 & 0\\
0 & 1 & 0 & 0
\end{array}
\right]  .
\end{equation}
The four vectors in (\ref{eq:ent-code-vecs}) determine an
entanglement-assisted code. We place them as row vectors in a parity check
matrix $H$:%
\begin{equation}
H=\left[
\begin{array}
[c]{cccc}%
1 & 1 & 0 & 1\\
-\sqrt{\frac{1}{2}} & \sqrt{2} & -\sqrt{2} & \sqrt{\frac{1}{2}}\\
1 & 0 & 1 & 0\\
-\sqrt{2} & \sqrt{\frac{1}{2}} & -\sqrt{\frac{9}{2}} & \sqrt{\frac{1}{2}}%
\end{array}
\right\vert \left.
\begin{array}
[c]{cccc}%
0 & 0 & 0 & 0\\
\sqrt{\frac{1}{2}} & -\sqrt{\frac{1}{2}} & \sqrt{\frac{1}{2}} & 0\\
0 & 1 & 0 & 0\\
\sqrt{\frac{1}{2}} & -\sqrt{2} & 0 & \sqrt{\frac{1}{2}}%
\end{array}
\right]  .\label{eq:ent-code-parity-check}%
\end{equation}
A symplectic matrix $\mathbf{\Upsilon}$\ relates $F$ to $H$. This symplectic
matrix $\mathbf{\Upsilon}$ determines the encoding circuit. We augment the
above matrices $F$ and $H$ to matrices $F_{\text{aug}}$ and $H_{\text{aug}}$
respectively. The augmented matrices $F_{\text{aug}}$ and $H_{\text{aug}}$
have the matrix $\left[
\begin{array}
[c]{cc}%
-I_{2\times2} & 0_{2\times2}%
\end{array}
\right]  ^{T}$ to the left of the vertical bar in $F$ and $H$ and the matrix
$\left[
\begin{array}
[c]{cc}%
0_{2\times2} & I_{2\times2}%
\end{array}
\right]  ^{T}$ as the last columns of $F$ and $H$ respectively. All the rows
in the augmented parity check matrices $F_{\text{aug}}$ and $H_{\text{aug}}$
are then orthogonal with respect to the symplectic product and therefore
correpond to a commuting set of observables via the map $\mathbf{M}$. We later
confirm that this code corrects for an arbitrary single-mode error.

Our main general result is as follows. There exists a continuous-variable
entanglement-assisted code with the following properties. Alice encodes her
state with the operation $U_{\mathbf{\Upsilon}}^{-1}U_{0}$. The set $S$ of
correctable errors obeys the following constraint:%
\begin{multline*}
\forall\ \mathbf{u},\mathbf{u}^{\prime}\in S:\mathbf{u}\neq\mathbf{u}^{\prime
},\\
\mathbf{u}-\mathbf{u}^{\prime}\notin C\ \ \ \vee\ \ \ \mathbf{u}%
-\mathbf{u}^{\prime}\in\text{iso}\left(  C^{\bot}\right)  .
\end{multline*}
The codespace $\mathcal{C}$ is the simultaneous zero eigenspace of the ordered
set:%
\begin{equation}
\mathcal{M}\equiv\left\{  \mathbf{M}\left(  \mathbf{h}\right)  :\mathbf{h}%
\text{ is a row of }H_{aug}\right\}  .
\end{equation}
Performing $U_{\mathbf{\Upsilon}}$, measuring the operators in $\mathcal{M}%
_{0}$ is equivalent to measuring operators in $\mathcal{M}$ followed by
performing $U_{\mathbf{\Upsilon}}$. Suppose an error $\mathbf{D}\left(
\mathbf{u}\right)  $ occurs where $\mathbf{u}\in S$. The general error set
relates to the canonical set by the mapping in Theorem~\ref{thm_unitary}:
$\left[  U_{\mathbf{\Upsilon}}\mathbf{D}\left(  \mathbf{u}\right)
U_{\mathbf{\Upsilon}}^{-1}\right]  =\left[  \mathbf{D}\left(  \mathbf{\Upsilon
u}\right)  \right]  $. Bob measures the reduced syndrome $\mathbf{r}\ $by
measuring the observables in the set $\mathcal{M}$. Bob finds the error
$\mathbf{u}$ corresponding to the reduced syndrome $\mathbf{r}$ and performs
$\mathbf{D}\left(  -\mathbf{u}\right)  $ to undo the error. Figure
\ref{fig:ea-code} illustrates the above operations for an
entanglement-assisted code.

The code corresponding to the parity check matrix in
(\ref{eq:ent-code-parity-check}) corrects for an arbitrary single-mode error.
Suppose that an error $D\left(  u\right)  $ occurs on the first mode. We set
$u=\left(  p|x\right)  $ and $p,x\in\mathbb{R}$ so that $p$ is a
momentum-quadrature error and $x$ is a position-quadrature error. Then Bob
measures the error syndrome to be as follows:%
\[%
\begin{bmatrix}
x & \sqrt{1/2}\left(  p-x\right)   & x & \sqrt{1/2}p-\sqrt{2}x
\end{bmatrix}
.
\]
Suppose the error $D\left(  u\right)  $ occurs on modes two, three, or four.
The error syndromes in respective order are then as follows:%
\begin{align*}
&
\begin{bmatrix}
x & \sqrt{2}x-\sqrt{1/2}p & p & \sqrt{1/2}x-\sqrt{2}p
\end{bmatrix}
,\\
&
\begin{bmatrix}
0 & -\sqrt{2}x+\sqrt{1/2}p & x & -\sqrt{9/2}x
\end{bmatrix}
,\\
&
\begin{bmatrix}
x & \sqrt{1/2}x & 0 & \sqrt{1/2}\left(  p+x\right)
\end{bmatrix}
.
\end{align*}
The above error syndromes are unique for any nonzero $p$ and $x$. Bob can
uniquely identify on which mode the error $D\left(  u\right)  $ occurs and
correct for it.

\section{Linear-Optical Encoding Algorithm}

We give an algorithm\ for decomposing an arbitrary encoding circuit into one
and two-mode operations using linear optics. The algorithm is an alternative
to the one given in \cite{braunstein:055801}. The unitary $U_{\mathbf{\Upsilon
}}^{-1}$ for the encoding circuit is an element of the group $\mathcal{G}%
_{n}^{Sp}$ that preserves the phase-free Heisenberg-Weyl\ group up to
conjugation \cite{PhysRevA.64.012310, PhysRevLett.88.097904}. The symplectic
group Sp$\left(  2n,\mathbb{R}\right)  $ is isomorphic to $\mathcal{G}%
_{n}^{Sp}$. Previous results show that any $\mathcal{G}_{n}^{Sp}$
transformation admits a decomposition in terms of linear optical elements and
squeezers \cite{braunstein:055801,PhysRevA.65.042304}. Our algorithm is a
different technique for determining the encoding unitary. It uses a symplectic
Gaussian elimination technique similar to a discrete-variable algorithm
\cite{hostens:042315}.

The Fourier transform gate, two-mode quantum nondemolition\ interactions, a
squeezer, and a continuous-variable\ phase gate generate all transformations
in $\mathcal{G}_{n}^{Sp}$. A position-quadrature squeezer $S_{i}\left(
a\right)  $ on mode $i$\ rescales the position quadrature\ by $a$ with
reciprocal scaling by $1/a$ in the momentum quadrature:%
\[
\hat{x}_{i}\rightarrow a\hat{x}_{i},\ \ \ \hat{p}_{i}\rightarrow\hat{p}%
_{i}/a.
\]
A Fourier transform $F_{i}$\ on mode $i$ acts as%
\[
\hat{x}_{i}\rightarrow-\hat{p}_{i},\ \ \ \hat{p}_{i}\rightarrow\hat{x}_{i}.
\]
A two-mode position-quadrature nondemolition interaction $Q_{12}^{X}\left(
g\right)  $\ with interaction strength $g$\ transforms the quadrature
observables as%
\begin{align*}
\hat{x}_{1}  &  \rightarrow\hat{x}_{1},\ \ \ \hat{p}_{1}\rightarrow\hat{p}%
_{1}-g\hat{p}_{2},\\
\hat{x}_{2}  &  \rightarrow\hat{x}_{2}+g\hat{x}_{1},\ \ \ \hat{p}%
_{2}\rightarrow\hat{p}_{2}.
\end{align*}
A two-mode momentum-quadrature nondemolition interaction\ $Q_{12}^{P}\left(
g\right)  $ with interaction strength $g$ transforms the quadrature
observables as%
\begin{align*}
\hat{x}_{1}  &  \rightarrow\hat{x}_{1}-g\hat{x}_{2},\ \ \ \hat{p}%
_{1}\rightarrow\hat{p}_{1},\\
\hat{x}_{2}  &  \rightarrow\hat{x}_{2},\ \ \ \hat{p}_{2}\rightarrow\hat{p}%
_{2}+g\hat{p}_{1}.
\end{align*}
A\ position-quadrature phase gate $P^{X}\left(  g\right)  $ with interaction
strength $g$ transforms the quadrature observables as%
\[
\hat{x}\rightarrow\hat{x},\ \ \ \hat{p}\rightarrow\hat{p}+g\hat{x},
\]
and a momentum-quadrature\ phase gate\ $P^{P}\left(  g\right)  $\ transforms
the quadrature observables as%
\[
\hat{x}\rightarrow\hat{x}+g\hat{p},\ \ \ \hat{p}\rightarrow\hat{p}.
\]
Filip et al. implemented $S\left(  a\right)  $, $Q_{12}^{X}\left(  g\right)
$, and $Q_{12}^{P}\left(  g\right)  $\ using linear optics \cite{filip:042308}.

We provide an implementation of the continuous-variable phase gate. Begin with
two modes---we wish to perform the phase gate on mode one. Suppose mode two is
a position-squeezed ancilla mode. Perform a position-quadrature nondemolition
interaction\ $Q_{12}^{X}\left(  g_{1}\right)  $ on modes one and two:%
\begin{align*}
\hat{x}_{1}  &  \rightarrow\hat{x}_{1},\ \ \ \hat{p}_{1}\rightarrow\hat{p}%
_{1}-g_{1}\hat{p}_{2},\\
\hat{x}_{2}  &  \rightarrow\hat{x}_{2}+g_{1}\hat{x}_{1},\ \ \ \hat{p}%
_{2}\rightarrow\hat{p}_{2}.
\end{align*}
Fourier transform mode two:%
\begin{align*}
\hat{x}_{1}  &  \rightarrow\hat{x}_{1},\\
\hat{p}_{1}-g_{1}\hat{p}_{2}  &  \rightarrow\hat{p}_{1}-g_{1}\hat{p}_{2},\\
\hat{x}_{2}+g_{1}\hat{x}_{1}  &  \rightarrow-\hat{p}_{2},\\
\hat{p}_{2}  &  \rightarrow\hat{x}_{2}+g_{1}\hat{x}_{1}.
\end{align*}
Perform a momentum-quadrature nondemolition interaction\ $Q_{12}^{P}\left(
g_{2}\right)  $ on modes one and two:%
\begin{align*}
\hat{x}_{1}  &  \rightarrow\hat{x}_{1},\\
\hat{p}_{1}-g_{1}\hat{p}_{2}  &  \rightarrow\hat{p}_{1}-g_{1}\hat{p}_{2}%
+g_{2}\left(  \hat{x}_{2}+g_{1}\hat{x}_{1}\right)  ,\\
-\hat{p}_{2}  &  \rightarrow-\hat{p}_{2}-g_{2}\hat{x}_{1},\\
\hat{x}_{2}+g_{1}\hat{x}_{1}  &  \rightarrow\hat{x}_{2}+g_{1}\hat{x}_{1}.
\end{align*}
Measure the position quadrature\ of mode two to get result $x$.\ Mode one
collapses as
\begin{align*}
\hat{x}_{1}  &  \rightarrow\hat{x}_{1},\\
\hat{p}_{1}-g_{1}\hat{p}_{2}+g_{2}\left(  \hat{x}_{2}+g_{1}\hat{x}_{1}\right)
&  \rightarrow\hat{p}_{1}+g_{1}x+g_{2}\hat{x}_{2}+2g_{2}g_{1}\hat{x}_{1}.
\end{align*}
Correct the momentum of mode 2 by displacing by $g_{1}x$ so that
\begin{align*}
\hat{x}_{1}  &  \rightarrow\hat{x}_{1},\\
\hat{p}_{1}+g_{1}x+g_{2}\hat{x}_{2}+2g_{2}g_{1}\hat{x}_{1}  &  \rightarrow
\hat{p}_{1}+g_{2}\hat{x}_{2}+2g_{2}g_{1}\hat{x}_{1}.
\end{align*}
The Heisenberg-picture quadrature observables for mode one are approximately
$\hat{x}_{1}$, $\hat{p}_{1}+2g_{2}g_{1}\hat{x}_{1}$ because the original
quadrature $\hat{x}_{2}$ has position-squeezing. So we implement a
continuous-variable\ position-quadrature\ phase gate $P^{X}\left(
g=2g_{2}g_{1}\right)  $.

We use the above gates to detail a symplectic Gaussian elimination procedure.
This procedure decomposes an arbitrary encoding circuit whose symplectic
matrix is $\mathbf{\Upsilon}$.

\begin{enumerate}
\item If $\mathbf{\Upsilon}_{1,1}$ equals zero, permute the first mode with
the second. Continuing permuting modes until $\mathbf{\Upsilon}_{1,1}$ is
nonzero. Normalize $\mathbf{\Upsilon}_{1,1}$ by simulating $S_{1}\left(
\mathbf{\Upsilon}_{1,1}^{-1}\right)  $.

\item Simulate $Q_{1i}^{X}\left(  -\mathbf{\Upsilon}_{i,1}\right)  $\ for all
$i\in\left\{  2,\ldots,n\right\}  $. The first column then has the form%
\[%
\begin{bmatrix}
1 & 0 & \cdots & 0 & \mathbf{\Upsilon}_{n+1,1} & \mathbf{\Upsilon}_{n+2,1} &
\cdots & \mathbf{\Upsilon}_{2n,1}%
\end{bmatrix}
^{T}.
\]

\item Simulate $P_{1}^{X}\left(  -\mathbf{\Upsilon}_{n+1,1}\right)  $ followed
by $F_{1}$.

\item Simulate $Q_{1i}^{P}\left(  -\mathbf{\Upsilon}_{j,1}\right)  $ for all
$i\in\left\{  2,\ldots,n\right\}  $ and $j=i+n$. Perform $F_{1}^{-1}$. The
first column has the form $%
\begin{bmatrix}
1 & 0 & \cdots & 0
\end{bmatrix}
^{T}$.

\item Name the new matrix $\mathbf{\Upsilon}^{\prime}$. Proceed to decouple
column $n+1$ of $\mathbf{\Upsilon}^{\prime}$. Matrix element $\mathbf{\Upsilon
}_{1,1}^{\prime}=1$ because $\mathbf{\Upsilon}^{\prime}$ is symplectic.
Simulate $Q_{1i}^{P}\left(  -\mathbf{\Upsilon}_{i+j,n+1}\right)  $ for all
$i\in\left\{  2,\ldots,n\right\}  $ and $j=i+n$.

\item Simulate $P_{1}^{P}\left(  -\mathbf{\Upsilon}_{1,n+1}\right)  $. Perform
$F_{1}^{-1}$.

\item Simulate $Q_{1i}^{X}\left(  -\mathbf{\Upsilon}_{i,1}\right)  $ for all
$i\in\left\{  2,\ldots,n\right\}  $. Perform $F_{1}$.
\end{enumerate}

The first round of the algorithm is complete and the new matrix
$\mathbf{\Upsilon}^{\prime\prime}$ has its first row and column equal to
$\mathbf{e}_{1}$, its $\left(  n+1\right)  ^{st}$ row and column equal to
$\mathbf{e}_{n+1}$, and all other entries equal to the corresponding entries
in $\mathbf{\Upsilon}$. The remaining rounds of the algorithm consist of
applying the same procedure to the submatrix formed from rows and columns
$2,\ldots n,n+2,\ldots,2n$ of $\mathbf{\Upsilon}$. All of the operations in
the algorithm consist of one and two-mode operations implementable with linear
optics. The encoding circuit is the inverse of all the operations put in
reverse order.

\section{Conclusion}

We have constructed a general theory of entanglement-assisted error correction
for continuous-variable quantum information. The theory of continuous-variable
quantum error correction broadens when Alice and Bob share a set of entangled
modes. They begin with a set of noncommuting observables that have good
error-correcting properties. They then employ shared entanglement to resolve
the anticommutativity in the original observables.

Our codes suffer from the same vulnerabilities as Braunstein's earlier codes
for continuous variables \cite{prl1998braunstein_error}. But the theory should
be useful as experimentalists improve the quality of squeezing and homodyne
detection technology.

Our example of a continuous-variable entanglement-assisted code requires two
entangled modes and corrects for an arbitrary single-mode error.

We also provided a way to construct encoding circuits using passive optical
elements, homodyne measurements, feedforward control, conditional
displacements, and off-line squeezers. The algorithm decomposes the encoding
circuit in terms of a polynomial number of gates. The algorithm requires a
large number of squeezers to implement an encoding circuit. But this scheme
for encoding should become feasible as technology improves.

\section{Acknowledgements}

The authors\ thank Igor Devetak for useful discussions and Min-Hsiu Hsieh for
Matlab code. MMW acknowledges support from NSF\ Grants CCF-054845\ and
CCF-0448658, and TAB and HK acknowledge support from NSF Grant CCF-0448658.

\bibliographystyle{apsrev}
\bibliography{eaqec-cv}

\end{document}